\documentclass[reqno,12pt]{amsart}
\usepackage{amsmath, latexsym, amsfonts, amssymb}
\usepackage{mathrsfs,enumerate}
\usepackage{graphics,epsfig,color}

\numberwithin{equation}{section}

\newcommand{\cG}{{\ensuremath{\mathcal G}} }
\newcommand{\cI}{{\ensuremath{\mathcal I}} }

\newfont{\indic}{bbmss12}

\newcommand{\hf}{{\frac{1}{2}}}

\newcommand{\bbN}{{\ensuremath{\mathbb N}} }

\newcommand{\bbP}{{\ensuremath{\mathbb P}} }

\newcommand{\myeqnarray}[1]{
  \begingroup
  \jot=#1pt
  \arraycolsep=2pt
  \begin{eqnarray}}

\newcommand{\eeqnarray}{\end{eqnarray}\endgroup}
\newcommand{\ba}{\begin{array}{cc}}
\newcommand{\ea}{\end{array}}

\begin{document}
\title{Macroscopic fluctuations theory of aerogel dynamics}
\author{Rapha\"el Lefevere}
\address{Laboratoire de Probabilit\'es et Mod\`eles Al\'eatoires\\
UFR de Math\'ematiques Universit\'e Paris 7
 Case 7012, 75205 Paris Cedex 13, France }
 \author{Mauro Mariani}
 \address{Laboratoire d'Analyse,
Topologie, Probabilit\'es (CNRS UMR 6632), Universit\'e Paul
C\'e\-zanne Aix-Marseille 3, Fa\-cult\'e des Sciences et Techniques
Saint-J\'er\^ome, Avenue Escadrille Normandie-Niemen 13397 Marseille
Cedex 20, France}
 \author{Lorenzo Zambotti}
 \address{Laboratoire de Probabilit{\'e}s
   et Mod\`eles Al\'eatoires (CNRS UMR. 7599)
 Universit\'e Paris 6 -- Pierre et Marie Curie,
U.F.R. Math\'ematiques, Case 188, 4 place
   Jussieu, 75252 Paris cedex 05, France}


\begin{abstract}
We consider the thermodynamic potential describing the macroscopic fluctuation of the current and local energy of a general class of Hamiltonian models including aerogels.  We argue that this potential is neither analytic nor strictly convex, a property that should be expected in general but missing from models studied in the literature.  This opens the possibility of describing  in terms of a thermodynamic potential  non-equilibrium phase transitions in a concrete physical context.
This special behaviour of the thermodynamic potential is caused by the fact that the energy current is carried by particles which may have arbitrary low speed with sufficiently large probability.

\end{abstract}


\maketitle
\section{Introduction}
In equilibrium statistical mechanics, the Boltzmann-Gibbs probability distribution is at the heart of the connection between the micro- and macroscopic level of systems made of a large number of particles $N$.  From that distribution, one may compute the relevant thermodynamic potentials which may be interpreted as rate functions at speed $N$ governing the large deviations of the macroscopic quantity of interest.  Typically, the lack of strict convexity of the thermodynamic potentials is the signature of the existence of phase transitions.  Out of equilibrium,  there is not a such thing as the Boltzmann-Gibbs formula for the microscopic configurations and, in order to understand the macroscopic behaviour of a given material, one must proceed differently.  Building on the seminal work of Onsager and Machlup \cite{OM}, several authors have built a general picture of the macroscopic current fluctuations in non-equilibrium statistical mechanics, see \cite{jona0,BD2} for overviews of the subject.  Although this approach is rigorously justified only in the framework of stochastic interacting particles, the scope is much more ambitious.   Macroscopic fluctuation theory describes, after a diffusive rescaling of space and time, the macroscopic behaviour of extended systems crossed by a flow of a conserved quantity: energy, mass or other relevant quantities. In this approach, the classical thermodynamical potentials are replaced by a rate function describing the fluctuations in space-time of the local currents and energies.   From the point of view of the theory of large deviations this rate functional is strictly analogous to the equilibrium thermodynamic potentials in the sense that it is a large deviations rate function at speed $N$.  From the knowledge of this rate function one may, as in equilibrium,  compute the cumulants of the relevant physical quantities. Starting from a quadratic form for the rate functionnal, this is what is achieved in \cite{BD1} where a relation for the successive cumulants of the current is obtained.  A fundamental issue, on which almost no result exist for realistic physical systems, is to determine the form of the non-equilibrium rate function.

In this paper, our goal is to derive the form that this rate function must take in a general class of Hamiltonian models including aerogels. These materials are gels whose liquid component has been removed and replaced by atoms of gases.  Because of their particular insulating properties, they have proved useful in  applications ranging from the manufacturing of new insulating glasses to the engineering of key parts of space probes.  The theoretical models we will consider are easy to study numerically and well described by a local equilibrium assumption \cite{GG,GL} in a weakly interacting regime.  Using a stochastic approximation, we show that the function describing the macroscopic fluctuations of those models is drastically different from the one generally expected and obtained since the work of Onsager and Machlup  \cite{jona1,jona2,jona0,jona3,BD1,BD2,BD3,OM}. As a function of the current it is {\it not} a quadratic function around the stationary current.  More than that, it is neither a smooth nor a strictly convex  function. This means, by analogy with the equilibrium case, that aerogels must exhibit interesting out of equilibrium phase transitions phenomena. 
This should also apply to a class of models made of tracers and scatterers recently considered by several authors \cite{ EckmannYoung, EckmannYoung2,Larralde,LinYoung,Mejia}.

We emphasize that our main working hypothesis is the validity of the description of  the systems at a diffusive space-time scale.  This amounts to assuming the {\it existence} of a rate function and does not have any implication {\it a priori}  on the specific shape that this function must take.  The validity of the diffusive description of macroscopic systems should hold for a large class of mechanical systems.  On the other hand, there are very few a priori constraints imposed on the rate function except a global one imposed by the Gallavotti-Cohen symmetry and arguably local ones (on the first two derivatives) imposed by linear response theory.  We will see, however, that the collisional mechanism by which the energy is transported through the system by the microscopic components is responsible for a singular shape of the rate function.  The two constraints stated above are still satisfied.

\section{The models}

 The dynamics that we want to consider is as follows.  
Consider $N$ particles  of unit mass with positions and momenta
$(\underline{\mathbf{q}}, \underline{\mathbf{p}}) \equiv
\big\{(\mathbf{q}_i, \mathbf{p}_i)\big\}_{1\leq i\leq N}$, with
$\mathbf{q}_i, \mathbf{p}_i \in \mathbb{R}^d$.  The positions are measured with respect to $N$ fixed centers located on a 1D lattice. The Hamiltonian $H$ takes
the form
\begin{equation}
H(\underline{\mathbf{p}}, \underline{\mathbf{q}})
= \sum_{i=1}^N \left[\frac{\mathbf{p}_i^2}{2} +V(\mathbf{q}_i)+
U(\mathbf{q}_{i}-\mathbf{q}_{i+1}) \right],
\label{Hamilton}
\end{equation}
where the interaction potential $U$ is equal to zero inside a region $\Omega_U\subset\mathbb{R}^d$ with
smooth boundary $\Lambda$ of dimension $d-1$, and equal to infinity outside.
Likewise, the pinning potential $V$ is assumed to be zero inside a
bounded region $\Omega_V$ and infinite outside, implying that the motion of
a single particle remains confined for all times. The regions $\Omega_U$ and
$\Omega_V$ being specified, the dynamics is equivalent to a billiard in
high dimension.  A typical example is given by the figure below.  Each disk moves freely within its square cell and collides with a neighbor when they both get sufficiently close to the hole located in the wall separating their adjacent cells.

\begin{figure}[thb]
\includegraphics[width =.99\textwidth]{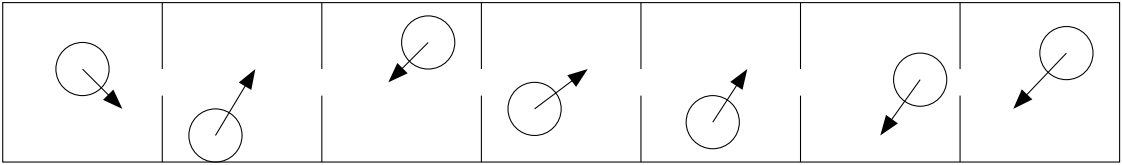} 
\caption{ Simplified aerogel dynamics}
\end{figure}
The first type of such models, in which the local dynamics is described by semi-dispersing billiards, was introduced in \cite{bunilive} and its thermal transport properties were studied in \cite{GG}.  General collisional models of this type have been introduced and analyzed in \cite{GL}.  To model those systems out of equilibrium, one couples them at their boundaries with stochastic thermal walls at different temperatures.  Rigorously, almost nothing is known about the invariant state of such systems.  Heuristically, however, the behavior of such systems is easy to guess \cite{GL,LefevereZambotti1}.
The evolution of the energy of the $n$-th particle may be written as,
\begin{equation}\label{evolen}
E_n(t)-E_n(0)=J_{n-1\to n}[0,t]-J_{n\to n+1}[0,t].
\end{equation}
where
\begin{eqnarray}
  J_{n\to n+1}[0,t]&=& \hf
  \sum_{0\leq k\leq N_t} \left[ p_{n}^\bot(s_k)^2 -
    p^\bot_{n+1}(s_k)^2 \right]\label{timeint},
 \end{eqnarray}
and $p^\bot_{n}$  is the component of the vector $\mathbf{p}_n$ in the direction of the unit vector
$\widehat{\mathbf{n}}=||q_i-q_{i+1}||^{-1}(q_i-q_{i+1})$ at the time of collision, namely $p^\bot_{n}=
\mathbf{p}_n\cdot \widehat{\mathbf{n}}$. The integer 
 $N_t$ counts the number of collisions up to time $t$ and $s_k$ is the $k$-th collision time.   Assume now that such systems  are thermalized at different temperatures at their boundaries.  In order to understand the transfer of energy from one side to the other, one is interested in the ergodic behavior of the current and in computing
$
\lim_{t\to +\infty}t^{-1}  J_{n\to n+1}[0,t],
$
giving the average current of energy in the stationary state.  Because of the special form of the time-integrated current (\ref{timeint}), a natural guess to make is that local equilibrium settles in and to assume that this limit is given by
\begin{equation}
\lim_{t\to +\infty}\frac{1}{t}  J_{n\to n+1}[0,t]=\nu \, (T_n-T_{n+1})
\label{Fourier}
\end{equation}
where $\nu=\lim_{t\to\infty} t^{-1}N_t$ is the frequency of collisions between neighbors under local equilibrium conditions and $T_n=\hf\langle p^2_n\rangle$ is the average kinetic energy of the particles.  The conductivity is thus identified to the frequency of collisions.  Typically, the collisions occur when the particles get near the boundaries of their cell and thus the frequency of collisions is roughly proportional to the  average number of visits to the boundaries per unit time. Because the particle travels freely within its cell, this is proportional to the square root of the average energy and thus to $\sqrt{T_n}$.  
 Numerical studies show  that the identification of the conductivity with the frequency of collisions holds true to a very high degree of accuracy in a wide class of collisional dynamics, when the individual particles collide rarely \cite{GG,GL}.

 \section{Macroscopic fluctuation theory}
 
 In macroscopic fluctuation theory,  the evolution of the relevant variables is described after a diffusive rescaling.  One introduces a macroscopic evolution time variable $s=t/N^2$ and a continuous space variable $x$ to describe functions on the spatial lattice by functions on the interval $[0,1]$. One assumes that the system may be described  by smooth functions $\hat J$ and $\epsilon$ related to the original time-integrated current and energy by
\begin{eqnarray}
J_{i\rightarrow i+1}[0,t]&=&N\hat J\left(\frac{i}{N},\frac{t}{N^2}\right)\nonumber\\
E_i(t)&=&\epsilon \left(\frac{i}{N},\frac{t}{N^2}\right).
\label{hydroscaling}
\end{eqnarray}
We define the instantaneous current at the macroscopic scale by
\[
j(x,s)=\partial_s \hat J(x,s)
\] 
and we introduce the notation:
$
j_k(s):=j\left(\frac k N,s \right).
$
The evolution of the energy is governed by (\ref{evolen}) which becomes at the macroscopic scale:
\begin{equation}
\partial_s \epsilon(x,s)=-\partial_x j(x,s).
\label{consen}
\end{equation}
 Macroscopic fluctuation theory predicts that asymptotically in $N$ for $j$ and $\epsilon$ related by (\ref{consen})
\begin{equation}
\bbP\left(\{\epsilon(x,s'),j(x,s')\}\right)\sim \exp[-N \,\hat\cI(j,\epsilon)] 
\label{ldN}
\end{equation}
where $\hat\cI(j,\epsilon)$ is given by
\begin{equation}
\hat \cI(j,\epsilon)=\int_0^s ds'\int_0^1 dx\; \phi(j(x,s'),\partial_x \epsilon(x,s'),\epsilon(x,s'))
\label{phi0}
\end{equation}
if $j$ and $\epsilon$ satisfy (\ref{consen}), and $\hat\cI=+\infty$ otherwise. The shape of $\phi$ is a priori quite arbitrary except for the constraints mentioned in the introduction.  Nevertheless, in the stochastic models  that have been studied so far, it is known that $\phi$ is a quadratic function of the current around the stationary current in the diffusive scaling limit  \cite{jona1,jona2,jona0,BD1,BD2,BD3}.  We assume that macroscopic fluctuation theory applies to our models in the sense that there exist smooth functions $\epsilon$ and $\hat J$ and that (\ref{ldN}) and (\ref{phi0}) hold true. 
We argue now that in the deterministic models described above $\phi$ can not be quadratic in $j$ and is rather given by (\ref{scaling}) below.
We assume  that the functions $\phi$, $j$ and $\epsilon$ are sufficiently smooth, we discretize space with increments of size $1/N$ and time with small steps of size $\Delta s$ and set:
\begin{eqnarray}
\hat \cI(j,\epsilon)
&=&\lim_{\Delta s\to 0}\lim_{N\to\infty}\frac{\Delta s}{N}\sum_{l=1}^{[s/\Delta s]}\sum_{k=1}^{N} \phi_{k,l}
\label{riemann}
\end{eqnarray}
With,  
$$
\phi_{k,l}:=
\phi\left (j_k(l\Delta s), N[E_{k+1}(l\Delta s)-E_{k}(l\Delta s)],E_k(l\Delta s)\right )
$$
and we identify $\exp\{-\Delta s\,\phi_{k,l}\}$ as the probability of observing an instantaneous current $j_k(l\Delta s)$ at site $k$ during a small macroscopic time interval $\Delta s$,  while the local energy and energy gradient are fixed. 
A key observation for our purpose is the following: as the macroscopic time varies by a small increment $\Delta s$, the energy $\epsilon$ changes only by an amount proportional to that increment.  On the other hand, in terms of the original microscopic variables, this corresponds to very long time scale $N^2\Delta s$.  Observing an instantaneous current $j_k(s)$ over a small interval of time  $\Delta s$ corresponds to observing an average time-integrated current
$
J_k[N^2 s,N^2(s+\Delta s)]=j_k(s)N\Delta s.
$
 Because of this relation
 we identify the probability of finding an instantaneous current $j_k(l\Delta s)$ with the probability of observing an average time-integrated current  (over a  microscopic time interval $N^2\Delta s$)
  equal to  $j_k(l\Delta s)/N$.  Thus,
 \begin{equation}
 \frac{ \phi_{k,l}}{N^2}=-\frac{1}{N^2\Delta s}\log\bbP\left(\frac{J_k[N^2l\Delta s,N^2(l+1)\Delta s)]}{N^2\Delta s}=\frac{j_k(l\Delta s)}{N}\right)
\label{phij}
\end{equation}
For $N$ large, the right-hand side converges to the large deviation functional of the time-integrated current $J$ evaluated at $j_k/N$, with the energy variables fixed. 
In our context, we are led to the following picture: a given particle makes a huge number of collisions with its neighbors while the average energy of its neighbors is basically unaffected.  We assume further that the dynamics is sufficiently chaotic so that the dynamical correlations  between neighbors are destroyed between two collisions among them.  This leads us to model the system during a time slice $\Delta s$ by the following stochastic dynamics.  Each particle moves freely in its cell and the effect of the collisions of a given particle with its neighbors is modeled by the action of hot walls at fixed temperatures.  The temperatures of the walls correspond to the average energies of the particles, which are fixed during the small macroscopic time interval $\Delta s$.
 Pictorially, the model of figure 1 is transformed into the model of the figure below.

\begin{figure}[thb]
\includegraphics[width =.99\textwidth]{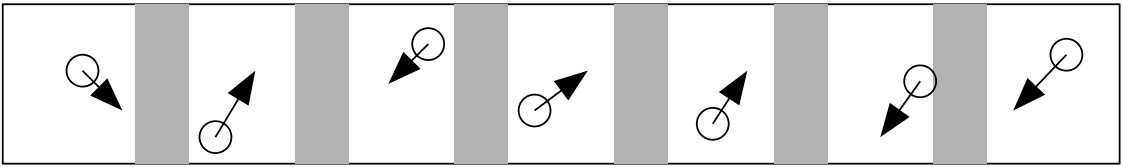} 
\caption{Collisions replaced by interactions with hot walls}
\end{figure}

Clearly, the direction orthogonal to the array of heat baths is irrelevant to the transport of energy and one is thus lead to the model of {\it confined tracers} introduced in \cite{LefevereZambotti1}.  Each particle moves freely in an interval and when it encounters the hot walls located at its boundaries, its velocity is reflected and it gets a new random speed $v$ distributed according to the distribution:
$$
\varphi_{\beta}(v)=\beta\, v \,e^{-\beta \frac{v^2}{2}}, \qquad v>0,
$$
where $\beta$ is the inverse of the temperature of the wall where the collision has taken place. We note that when the temperatures on the boundaries are equal, the invariant distribution of the random process describing the evolution of the position and velocity of the particle is given by the equilibrium  Maxwell-Boltzmann-Gibbs formula \cite{LefevereZambotti1}.  It is very important to note that if one takes as the updating rule a Maxwellian distribution, one does {\it not} get the Maxwell-Boltzmann-Gibbs distribution as the invariant measure for the random process \footnote{In the context of the study of gases contained in a vessel, it is customary to model the action of thermal walls by imposing special boundary conditions. We show in \cite{LMZ1} that this amounts to specify the domain of the generator of the random process corresponding to the above updating rule.}. In the stochastic model the energy current corresponding to a given particle is given by:
$$
J[0,t]=\sigma_0\frac 1 2\sum_{k=1}^{N_t}(-1)^k v_k^2
$$
where $v_k$ is the speed the particle gets at the $k$-th collision with a wall, $N_t$ is the number of collisions up to time $t$ and $\sigma_0$ is the sign of the initial velocity.  It is important to note at this stage that the form of the current is based on the deterministic part of dynamics, it simply corresponds to the kinetic energy carried by particles moving back and forth. As we will see below, $N_t$, whose large deviation properties will be central to our argument, depends in a simple fashion on the sequence of velocities assumed by the particle.  This is because the deterministic motion of the particle between the interactions with the walls is ballistic. One can show that,
$$
\lim_{t\to\infty}\frac{J[0,t]}{t}=\frac{T_L-T_R}{(\frac{\pi}{2T_L})^\hf+(\frac{\pi}{2T_R})^\hf}
$$
where $T_L$ and $T_R$ are the left and right temperatures.

 In  \cite{LMZ1}, we study in details the large deviations functional of the current $\cI(j,\tau,T)$ defined as:  
$$
\bbP_{\tau,T}\left (\frac{J[0,t]}{t}= j\right)\sim e^{-t \cI(j,\tau,T)},\quad t\to\infty.
$$
$\bbP_{\tau,T}$ is the probability associated to the stochastic dynamics with a fixed temperature difference $\tau=T_L-T_R$ between the two walls and average temperature $T=\frac{T_L+T_R}{2}$. 
More precisely, we show in \cite{LMZ1} the following result: if $\tau\ne0$ then, 
\begin{equation}
\lim_{\varepsilon\downarrow 0}\varepsilon^{-2}\cI(\varepsilon j,\varepsilon\tau,T)=\left\{\begin{array}{ll}
\frac{(j-\kappa\tau)^2}{4\kappa T^2} \ \ {\rm if} \ \ j\tau>\kappa\tau^2  \\
0  \ \ {\rm if} \ \ j\tau\in[0,\kappa\tau^2] 
\\ 
-\frac{j\tau}{2 T^2} \ \ {\rm if} \ \ j\tau\in[-\kappa\tau^2,0] 
\\
\\
\frac{j^2+\kappa^2\tau^2}{4\kappa T^2} \ \ {\rm if} \ \ j\tau<-\kappa\tau^2,
\end{array}
\right.
\label{scaling}
\end{equation}
\noindent  where $\kappa=(\frac{T}{2\pi})^\hf$.  When $\tau=0$, $\lim_{\varepsilon\downarrow 0}\varepsilon^{-2}\cI(\varepsilon j,\varepsilon\tau,T)=\frac{j^2}{4\kappa T^2}$.
Denoting the RHS by $\cG(j,\tau,T)$, we observe that it has the Gallavotti-Cohen symmetry:
$$
\cG(j,\tau,T)-\cG(-j,\tau,T)=\frac{j \tau}{2 T^2}.
$$
 The most striking feature of the functionals $\cI(\cdot,\tau,T)$ and $\cG(\cdot,\tau,T)$  is the fact that they vanish on the interval $[0,\kappa\tau]$.  The origin of this phenomenon may be traced back to the occurrence of slow velocities with sufficiently large probability in the course of the motion of the particle.  More precisely, the process can be entirely characterized by the sequence $(t_k)_{k\in\bbN}$ of waiting times between collisions of the particle with the walls.   Those waiting times are equal to the inverse of the random speeds $t_k=1/v_k$ and are thus distributed according to a density:
 $$
 \psi_\beta(u)=\frac{\beta}{u^3}e^{-\frac{\beta}{2u^2}}, \qquad u>0.
 $$
 The study of the large deviations of the current boils down to analyzing the large deviations of the Markov renewal process $N_t=\sup \{n:S_n\leq t\}$, where $S_0=0$ and $S_n=t_1+\ldots+t_n$.
The fact that the distribution $\psi_\beta(u)$ decays only polynomially as $u\to\infty$, implies that the large deviations functional of $N_t/t$ is identically equal to zero between $0$ and $\nu:=\lim_{t\to\infty} N_t/t$. A key point to see this is the following: for $\alpha\geq 0$,
\begin{eqnarray}
\bbP_{\tau,T}\left (\frac{N_t}{t}\leq \alpha\right)&=&\bbP_{\tau,T}\left (S_{[\alpha t]+1}>t\right)\nonumber\\
&\geq &\bbP_{\tau,T}\left(t_1> t\right)\sim t^{-2}
\end{eqnarray}
and taking the log on both sides of the inequalities, dividing by $t$, this implies that the large deviations functional vanishes at least in $\alpha=0$.  Taking care of technicalities which involves the proof of the convexity of the functional (see \cite{LMZ1}), one concludes.

We go back to the computation $\phi_{k,l}$: we basically identify the LHS of (\ref{phij}) with the large deviations functional of the stochastic dynamics. Namely, we write:

\begin{eqnarray}
\frac{\phi_{k,l}}{N^2}&=& \cI\left(\frac {j_k(l\Delta s)}{N},E_{k+1}(l\Delta s)-E_{k}(l\Delta s),E_k(l\Delta s)\right)\nonumber\\
&&+\, o\left(\frac 1 {N^2}\right).
\end{eqnarray}
Then, using (\ref{scaling}) and (\ref{riemann}) one obtains finally:
\begin{equation}
\hat \cI(j,\epsilon)=\int_0^s d\tau\int_0^1 dx\; \phi(j(x,s'),\partial_x \epsilon(x,s'),\epsilon(x,s'))
\end{equation}
where $\phi(j,\tau,T)$ is given by the RHS of (\ref{scaling}).

\begin{figure}[thb]
\includegraphics[width =.8\textwidth]{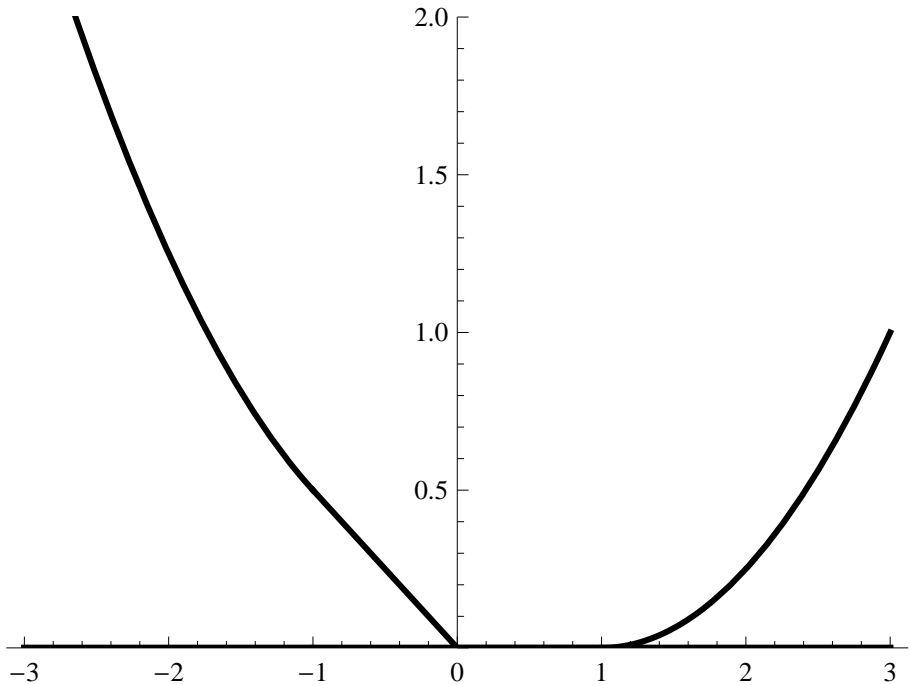} 
\caption{ Plot of $\phi$ as a function of $j$ for $\kappa\tau=\kappa T^2=1$}
\end{figure}
\section{Conclusions}
As a rigorous derivation of the relation (\ref{scaling}) for  deterministic systems seems at the moment out of reach, it is of great importance to derive observable consequences which could be tested either in physical experiments or in numerical simulations.  By analogy with the equilibrium case, we expect the lack of analyticity of $\phi$ to be responsible for the occurrence of phase transitions. How those will manifest themselves in physical or numerical experiments is at this stage an open question.  We note that those phase transitions are of different kind than the ones found in \cite{jona1,jona2,jona3,BD3}.  In those works, they are obtained from a quadratic $\phi$ when optimizing the energy profile to get a given current.  In our case, they are present in the system from the beginning, in the sense that $\phi$  itself is a non-analytic function.  It would be interesting to determine what happens to the relation giving the successive cumulants of the current derived in \cite{BD1} which was based on a quadratic form of $\phi$.  We expect the relation (\ref{scaling}) to be generic in local collisional dynamics, not only in the models we described above but also in deterministic models made of tracer particles and fixed scatterers  \cite{ EckmannYoung, EckmannYoung2,Larralde,LinYoung,Mejia}.  This is  because our argument is based only on the fact that the current is carried by a particle travelling ballistically between one collision and the next and on a local equilibrium hypothesis implying the occurence of particles with arbitrarily low speed in the system.  More precisely (\ref{scaling}) should govern the macroscopic fluctuations of the system whenever it appears to be well described by a local equilibrium distribution.  This applies to the dynamics of \cite{bunilive,GG,GL} in a weakly interacting regime and to the {\it confined} tracers models of  \cite{ EckmannYoung, EckmannYoung2,Larralde,LinYoung,Mejia}.  However, it does {\it not} apply to those among the models of  \cite{ EckmannYoung, EckmannYoung2,Larralde,LinYoung,Mejia} in which the tracers are allowed to travel from cell to cell. In those models,  the average over long times of the local transfer of energy can not be a local quantity.  Indeed the local transfer of energy is due to particles travelling all along the system and their average frequency of visits to a given cell will depend on the temperatures of every scatterers they meet during their journey.  
 We do not expect the behaviour described by (\ref{scaling}) to occur in other types of lattice Hamiltonian dynamics in which the components interact through a smooth potential.  We emphasize also that it is not the fact that we replaced the deterministic collisions by stochastic ones that is responsible for the special behaviour of $\phi$. This is caused by what is left of the deterministic dynamics, namely the ballistic motion of the particle. In particular, one should not expect the same phenomenon to occur in the case of self-consistent harmonic chain \cite{Bonetto,Bolsterli, Olla}.  The main striking feature of $\phi$, namely its flat part, is  due to the deterministic part that we have not modified. Therefore, one should expect this feature to be robust and independent of the approximations that we made for the collisional part of the dynamics.
We note finally that a similar large deviations functional has been found in the context of random walks in random environments \cite{Comets,Greven}.

\end{document}